\documentstyle[psfig]{kapproc} 

\kluwerbib

\startauthorindex

\newcommand{\ltsima}{$\; \buildrel < \over \sim \;$}
\newcommand{\simlt}{\lower.5ex\hbox{\ltsima}}
\newcommand{\gtsima}{$\; \buildrel > \over \sim \;$}
\newcommand{\simgt}{\lower.5ex\hbox{\gtsima}}

\def\pp{\noindent \parshape 2 5.9truecm 6truecm 5.9truecm 6truecm}

\def\gsim{ \lower .75ex \hbox{$\sim$} \llap{\raise .27ex \hbox{$>$} } }
\def\lsim{ \lower .75ex \hbox{$\sim$} \llap{\raise .27ex \hbox{$<$} } }

\def\happy{ \Large \lower 0.9ex \hbox{${\stackrel {\stackrel{\odot\odot}{o}} {\smile}}$} }
\def\sad{ \Large \lower 0.9ex \hbox{${\stackrel {\stackrel{\odot\odot}{o}} {\frown}}$} }
\def\sadish{ \Large \lower 0.9ex \hbox{${\stackrel {\stackrel{\odot\odot}{o}} {-}}$} }

\begin{document}

\articletitle{The IMF Long Ago and Far Away:}

\articlesubtitle{Faint Stars in the Ursa Minor dSph Galaxy}

\author{Rosemary F.G.~Wyse}
\affil{Johns Hopkins University\\
Department of Physics \& Astronomy}
\email{wyse@pha.jhu.edu}

\chaptitlerunninghead{The IMF Long Ago}

\anxx{R.F.G.~Wyse}

\begin{abstract}
The dwarf spheroidal galaxy in Ursa Minor is apparently dark-matter
dominated, and is of very low surface brightness, with total
luminosity only equal to that of a globular cluster.  Indeed its
dominant stellar population is old and metal-poor, very similar to
that of a classical halo globular cluster in the Milky Way Galaxy.
However, the environment in which its stars formed was clearly
different from that in the globular clusters in the Milky Way Galaxy
-- what was the stellar IMF in this external galaxy a long time ago?
The fossil record of long-lived, low-mass stars contains the
luminosity function, derivable from simple star counts.  This is
presented here.  The mass function requires a robust mass-luminosity
relation, and we describe the initial results to determine this, from
our survey for eclipsing low-mass binaries in old open clusters.  The
massive star IMF at early times is constrained by elemental abundances
in low-mass stars, and we discuss the available data.  All data are
consistent with an invariant IMF, most probably of Salpeter slope at
the massive end, with a turnover at lower masses.

\end{abstract}

\section*{Introduction}

The stellar IMF at high redshift is of great importance for a wide
range of astrophysical problems, such as the ionization and enrichment
of the intergalactic medium, the extragalactic background light, the
visibility of galaxies and the rate at which baryons are locked-up
into stars and stellar remnants.  There are two complementary
approaches to the determination of the IMF long ago and far away: one
is to observe directly high redshift objects, and attempt inferences
on the stellar IMF from the integrated spectrum and photometry, while
the second approach analyses the fossil record in old stars at low
redshift. The characterization of the stellar IMF in external
galaxies, compared to that in the Milky Way, is a crucial step in
deciphering the important physical processes that determine the
distribution of stellar masses under a range of different physical
conditions.  The low mass stellar IMF at the high redshifts at which
these stars formed is directly accessible through star counts, plus a
mass-luminosity relation.  The high mass IMF at these high redshifts
is constrained by the chemical signatures in the low mass stars that
were enriched by the supernovae from the high mass stars. 
I will discuss  both ends of the IMF at high redshift, in an external galaxy.

\section{Extremely Old Stars }

Simulations of galaxy formation within the framework of the
`concordance' ($\Lambda$)CDM cosmology agree that the first stars form
within structures that are less massive than a typical L$_\ast$ galaxy
today (e.g.~Kauffmann, White \& Guiderdoni 1993; Cole et
al.~2000).  Large galaxies  form hierarchically, through the
merging and assimilation of such smaller systems.  Satellite galaxies
of the Milky Way are survivors of this merging (e.g.~Bullock, Kravtsov
\& Weinberg ~2000). The stars that formed at early
times are found, at the present day, throughout large galaxies, and
also in satellite galaxies.  Environments with little subsequent star
formation are the best places to find and study old stars -- the
stellar halo of the Milky Way, and a few of the dwarf spheroidal
satellite galaxies.

\subsection{The Ursa Minor Dwarf Spheroidal Galaxy}

The dwarf spheroidal galaxy in Ursa Minor (UMi dSph), like all members
of its morphological class (Gallagher \& Wyse 1994), has extremely low
surface brightness, with a central value of only $\sim
25.5$~V~mag/sq.~arcsec, or $\sim 2.5\, { L}_\odot$/pc$^{-2}$.  The total luminosity is in the range $2-4 \times
10^5\,{ L}_{V,\odot}$ (Kleyna et al.~1998; Palma et al.~2003), equal to
that of a luminous Galactic globular cluster.  Again similar to a
globular cluster, the Ursa Minor dSph 
contains  little or no gas 
and has apparently not formed a significant
number of stars for $\sim 12$~Gyr (e.g.~Hernandez et al.~2000; 
Carrera et al.~2002), or since a redshift $z \simgt 2$.  The metallicity
distribution of the stars is narrow, with a mean of [Fe/H] $\sim
-1.9$~dex and a dispersion of $\sim 0.1$~dex (e.g.~Bellazzini
et al.~2002).   The stellar line-of-sight velocity dispersion is $\sim
10$~km/s (e.g.~Wilkinson et al.~2004), sufficiently large that, unlike
globular clusters, 
equilibrium models have a large mass-to-light ratio, $({\cal M}/L)_V \simgt
50~({\cal M}/L)_{V,\odot},$ perhaps as high as several hundred in solar units
if the mass is $\simgt 10^8~{\cal M}_\odot$ (Wilkinson et al.~2004),  
indicating a non-baryonic dark halo.
Non-equilibrium models are rather contrived and themselves fail to
explain the data (e.g.~Wilkinson et al.~2004).  Models of the
evolution of dwarf spheroidals are by no means well developed, but
very likely the stars formed in an environment rather different than
that of globular cluster stars, or of current star-forming regions in
the disk of the Milky Way.

The distance of the Ursa Minor dSph is only $\sim 70$~kpc, 
close enough that a determination of
the luminosity function of low-mass main sequence stars through star
counts is feasible, particularly using the Hubble Space Telescope.
The (unusually) simple stellar population of this dwarf spheroidal --
essentially of single age, single
metallicity --  makes the derivation of the luminosity function from star
counts a robust procedure.  This determines the low-mass stellar IMF
at redshifts of $\simgt 2$.  High-resolution spectroscopy of the
luminous evolved  stars is also possible, yielding elemental abundances
which constrain the high-mass stellar IMF  that enriched
the low-mass stars we observe. 

\section{The Faint Stellar Luminosity Function and Mass Function at Redshift $\simgt 2$}

The (very) dominant stellar population in the UMi dSph is very similar
to that of a classical Galactic halo globular cluster.  The most
robust constraint on the low-mass stellar IMF is then obtained by a
direct comparison between the faint stellar luminosity functions of
the UMi dSph and of representative globular clusters of the same age
and metallicity, such as M15 and M92, observed in the same bandpasses,
same telescope and detector. With the same stellar populations, this
is a comparison between {\it mass\/} functions, and differences may be
ascribed to variations in the low-mass IMF. 

We therefore obtained deep images with the Hubble Space Telescope in a
field close to the center of the UMi dSph, using STIS as the primary
instrument (optical Long Pass filter), with WFPC2 (V$_{606}$ and
I$_{814}$ filters) and NICMOS (NIC2/H-band) in parallel.  The WFPC2
filters matched those of extant data for M15 and M92; we obtained our
own 
STIS/LP and NIC2/H-band data for M15.  Similarly
exposed data for an \lq off\rq \/ field, at 2--3 tidal radii from the
centre of UMi dSph, were also acquired.  The detailed paper presenting the results from
the full dataset is Wyse et al.~(2002); preliminary results from 
a partial 
WFPC2 dataset were presented in Feltzing, Wyse \& Gilmore (1999). 

The images are not crowded and standard photometric techniques were
used to derive the luminosity functions.  For the WFPC2 data, the
luminosity functions were based only upon stars (unresolved objects)
that lie close to the well-defined UMi dSph main sequence locus in the
colour-magnitude diagram (CMD; see Figure~1).  The \lq off\rq\/ field CMD confirmed
little contamination from Galactic stars.  The STIS luminosity
function was based on one band only, and we employed various
approaches to background subtraction. The NICMOS data served only to
exclude a hypothetical population of very red stars and will not be
discussed further here.  The extant WFPC2 data for the globular
clusters (Piotto et al.~1997) are from fields at intermediate radii
within the clusters, where effects of dynamical evolution on the mass
function should be minimal. 
\begin{figure}[ht]
\vskip -0.25cm
\psfig{file=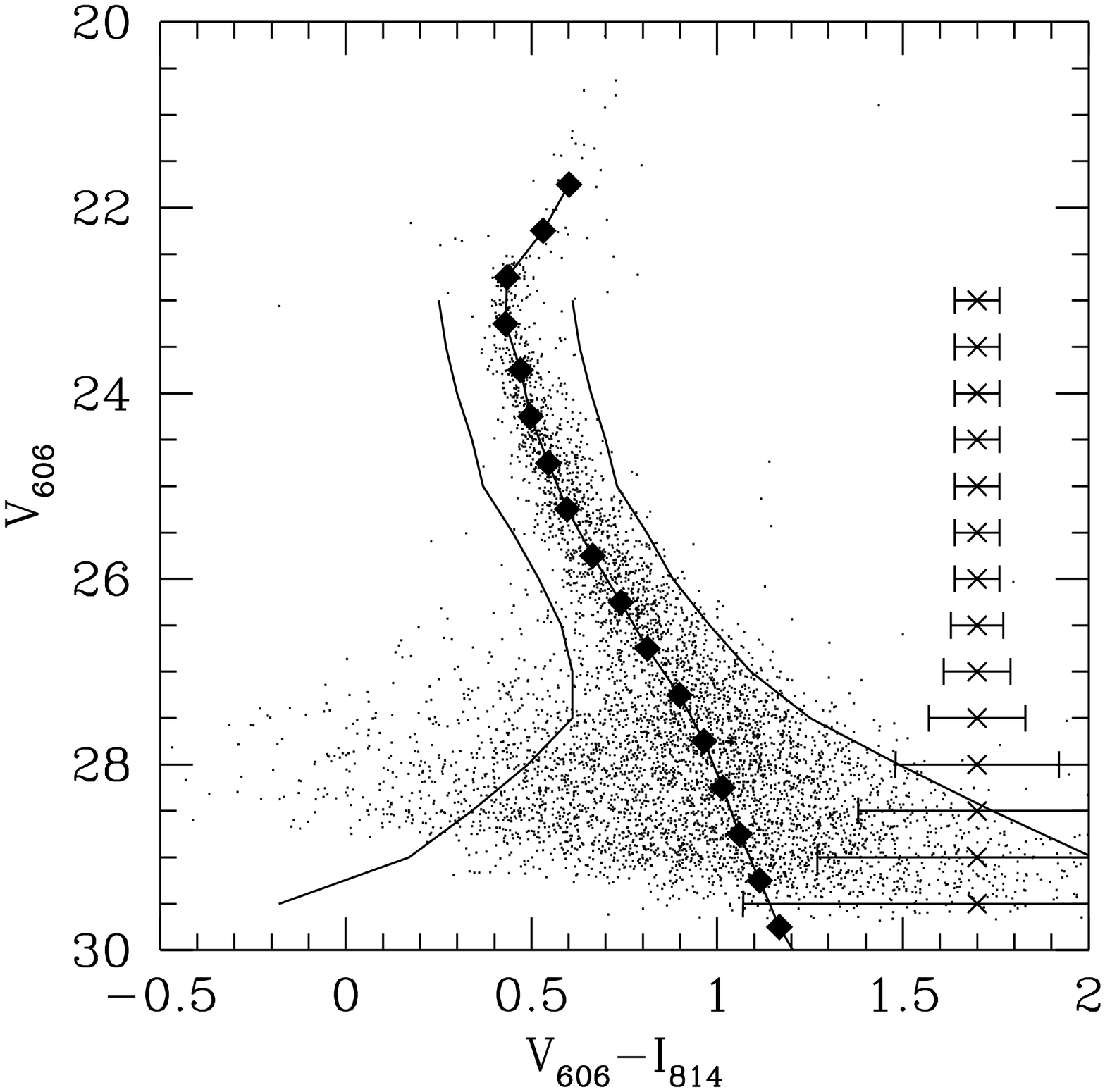,width=2.3in,angle=0}
\vskip -7.cm
\narrowcaption{} \par\pp {\small From Wyse et al.~(2002). 
Colour-magnitude diagram for all UMi dSph stars from the three
wide-field cameras on WFPC2. The full curves delineate the selection
criteria for stars to be included in the luminosity
function. The error-bars in each magnitude bin are shown at the right.
The well-defined narrow main sequence is the main feature. The few
blue stars close to the turn-off are probably blue stragglers, rather
than younger stars. The distribution of stars across the main sequence 
is asymmetric, to the red, and is consistent with
a normal population of binary stars.} 
\vskip -1cm
\end{figure}

\phantom{}

\subsection{Comparisons with Globular Clusters}
The comparisons with the WFPC2 colour-magnitude based V-band and
I-band luminosity functions are shown in Figure~2. We adopted 0.5~mag
bins to have reasonable numbers in each bin, and to minimize effects
of e.g.~reddening and distance moduli uncertainties. The 50\%
completeness limits for the UMi dSph data are equivalent to absolute
magnitudes of $M_{606} = +9.1$  and $M_{814} = +8.1$, which using the
Baraffe et al.~(1997) models {\it both\/}  
correspond to masses of ${\cal M} \sim 0.3\, {\cal M}_\odot $. The
STIS/LP data provide an independent check, by both a direct
LP-luminosity function comparison between M15 and UMi dSph, and a
derived STIS-based I-band luminosity function.  All show that the
globular cluster stars and the UMi stars have 
indistinguishable faint luminosity functions, down to an equivalent mass
limit of $\sim 0.3\, {\cal M}_\odot$ (see Wyse et al.~2002 for
details).   

\begin{figure}[ht]
\psfig{file=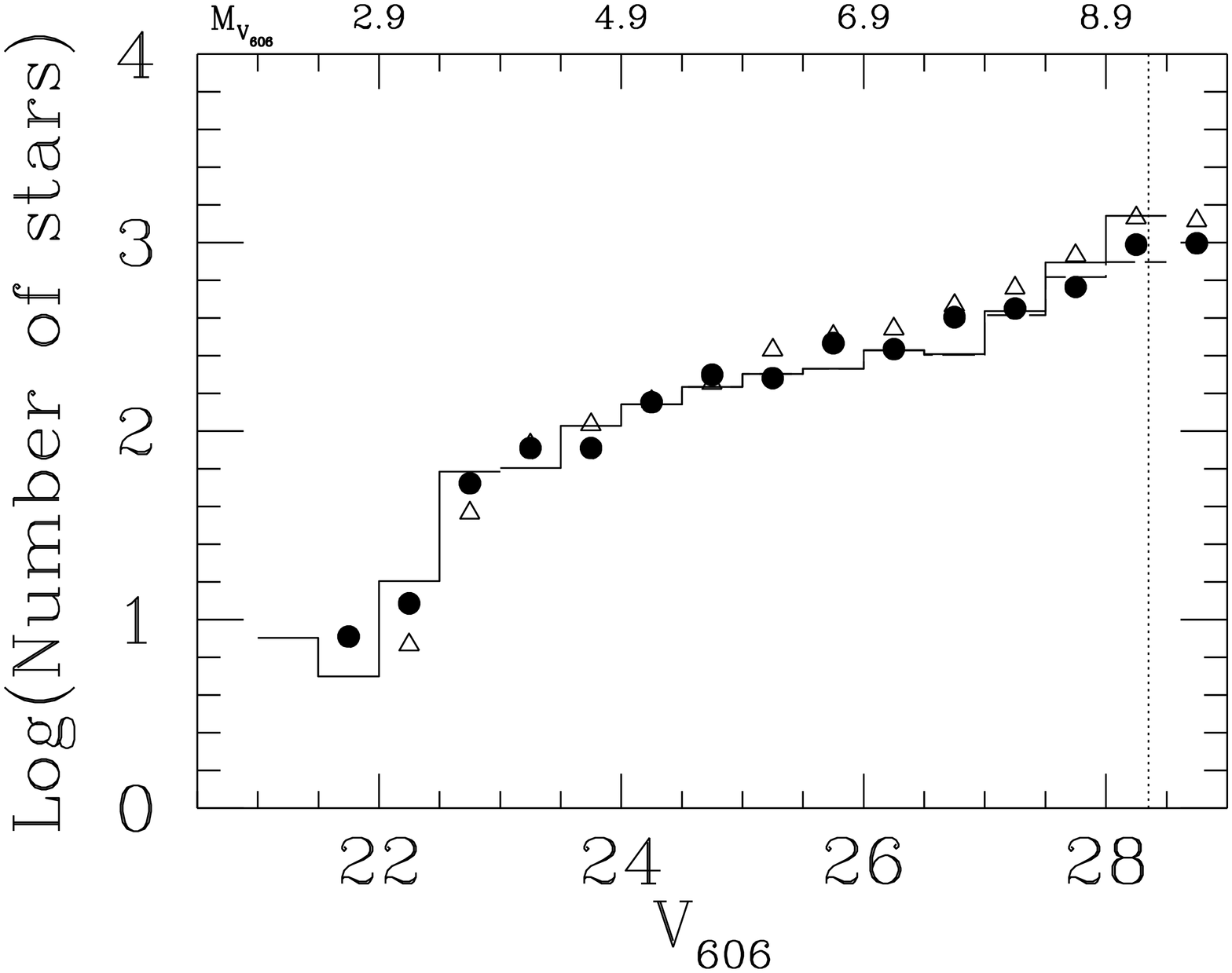,width=2.4in,angle=270}
\vskip -4.51cm
\hskip 2.32in
\psfig{file=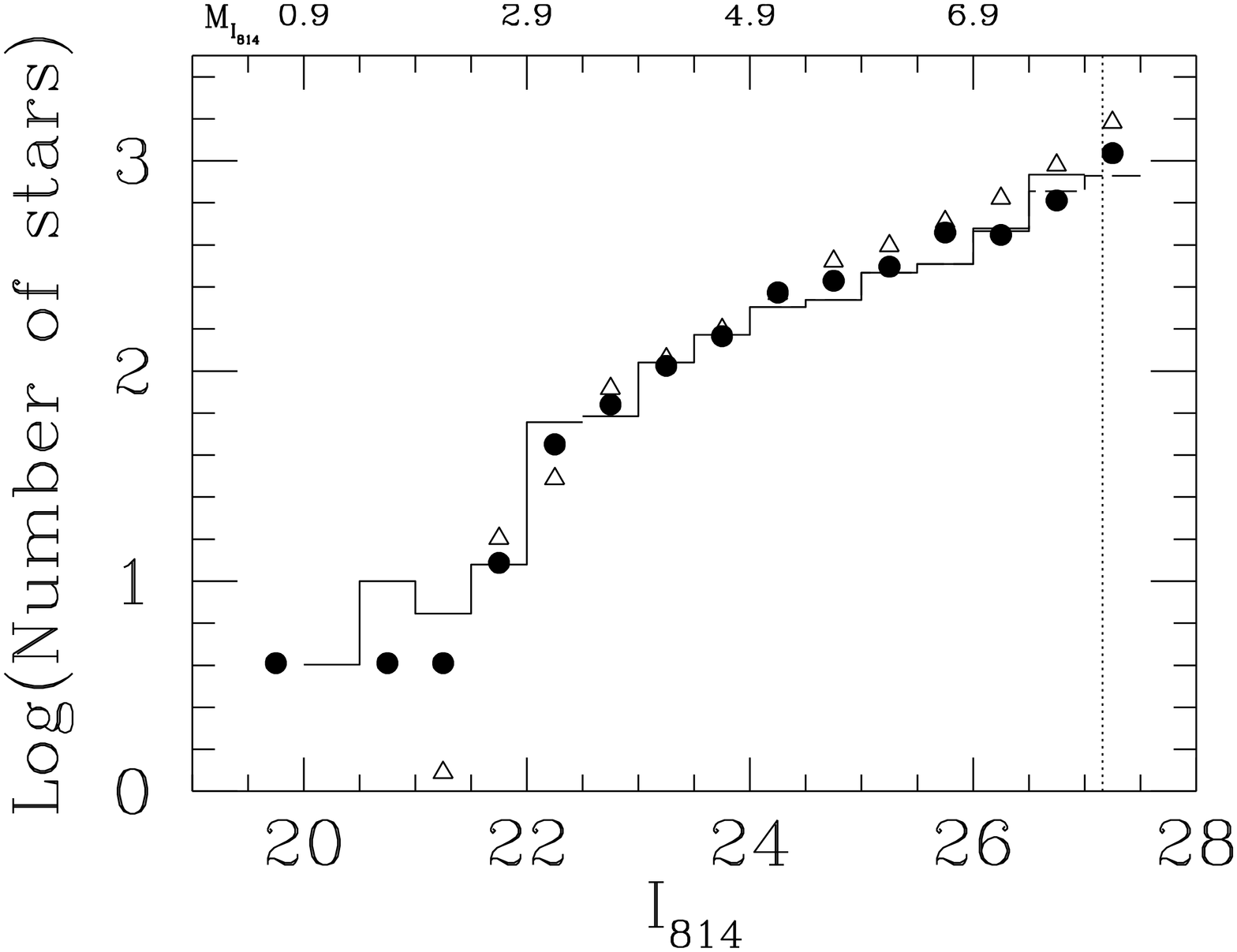,width=2.4in,angle=270}

\vskip -0.25truecm
\caption{Based on figures in Wyse et al.~(2002). Comparisons between the completeness-corrected Ursa Minor
luminosity functions (histograms; 
50\% completeness indicated by the vertical dotted line) 
in the V-band (left panel) and the I-band (right panel)  and the same 
for M92 (filled circles) and M15 (open triangles) (both taken from
Piotto et al.~1997, renormalized and shifted to the same distance as the Ursa Minor dSph). 
The luminosity functions for the 
globular clusters and the dwarf spheroidal galaxy are indistinguishable.}
\vskip -0.5cm
\end{figure}

We employed various statistical tests to quantify the agreement of the
various datasets -- e.g.~STIS-derived I-band {\it vs.}~WFPC2 I-band etc: 

$\heartsuit$ Linear, least-square fits to the (log) counts as a
function of apparent magnitude, using various ranges of magnitude and
differing bin choices, consistently found agreement to better than
2$\sigma$. 

$\heartsuit$ Kolmogorov-Smirnov tests on the unbinned data
for  a variety of magnitude ranges; the results depend on systematics
such as the relative distance moduli, but again there is general
agreement to better than 5\% significance level.

$\heartsuit$ $\chi-$square tests were carried out on the binned data,
using a variety of bin centers (maintaining 0.5~mag bin widths) and magnitude ranges
and again agreement to better than 5\% significance level.

The main result is that the underlying mass functions of low-mass
stars in Galactic halo globular clusters and in the external galaxy
the UMi dSph are indistinguishable.  This is a comparison between two
different galaxies, and systems of very different baryonic densities
and dark matter content.

\section{Low-Mass Stellar Mass Functions}

Adopting the Baraffe et al.~(1997) models, the 50\% completeness
limits for the luminosity functions of the stars in the UMi dSph
correspond to $\sim 0.3\, {\cal M}_\odot$, and the mass function may
be fit by a power law, with slope somewhat flatter than the Salpeter
(1955) value, over the range we test, of $0.3 \simlt {\cal M/M}_\odot \simlt 0.8$.
This is consistent with the solar neighbourhood mass function over
this mass range, and indeed the universal mass function that appears
to be the conclusion of this meeting.

However, the light-to-mass transformation is not robustly defined for
K/M dwarfs, especially as a function of age and metallicity.
Calibration of this is best achieved by analysis of low-mass stars in
detached eclipsing binary systems, and we have recently undertaken a
photometric 
survey of open clusters to identify candidate low-mass binary systems
to be followed up with spectroscopy for radial velocity curves;
this forms the PhD thesis of Leslie Hebb at Johns Hopkins University.

Our sample consists of six open clusters of known age and metallicity
(from the brighter turn-off stars), old enough to have low-mass stars
on the main sequence, age $\simgt 2 \times 10^8$~yr, with the oldest
being $\sim 4$~Gyr. We used both the
Wide Field Camera on the 2.5m Isaac Newton Telescope and the Mosaic Camera
on the Kitt Peak 4m telescope, each of which provide a field of view
of $\sim 35^\prime \times 35^\prime$.  The observing strategy we
adopted was designed to enable the detection of a 0.05~mag amplitude
eclipse in a target 0.3~${\cal M}_\odot$ star, monitored on timescales
of fraction of an hour, hours and days.  The low probability of
eclipse means that populous clusters must be observed for many
days. We expect our survey to find 3--5 low-mass eclipsing systems.
The details of the survey are presented in Hebb, Wyse \& Gilmore (2004). 

\begin{figure}[ht]
\centerline{\psfig{file=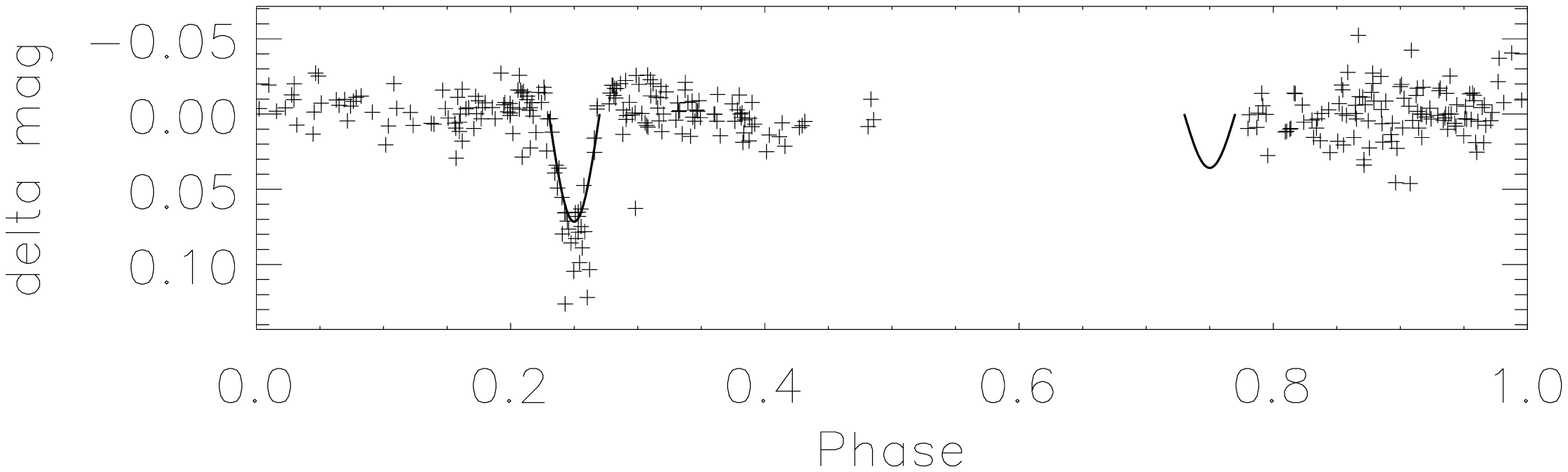,width=4.5in,angle=90}}

\caption{The phase-folded lightcurve for a candidate M-dwarf eclipsing
system in the open cluster M35, with a period of just over 1 day.
 The crosses show all differential photometry measurements for this object collected in January
 2002, January 2003 and February 2003.  The solid line is a simple sine-wave fit to the measurements
taken during the primary eclipse.
The phase of the secondary eclipse is marked by the same sine function fit
to the primary eclipse, but with half the amplitude.}

\end{figure}

The imaging data has all been acquired, differential photometry
obtained  and we are now analyzing
the derived light curves.  An example of the light curve of a
candidate eclipsing low-mass system is shown in Figure~3; the
candidate was identified by applying a box-fitting algorithm to the
photometric time series.  Our photometry, plus infrared data
from 2MASS, is consistent with this being an M-dwarf system.  We have 
applied for follow-up spectroscopic data, together with
higher time-sampling photometric data, for this system and for our
other candidates.

\section{High-Mass Stellar Mass Functions}

The high-mass stellar mass function long ago can be constrained by the
elemental abundances in the long-lived low-mass stars that formed from
gas that was enriched by the Type~II supernovae from the massive stars
of a previous generation (see e.g.~review of Wyse 1998).
Interpretation of the pattern of elemental abundances is easiest for
low-mass stars that formed early in a star-formation event, and were
enriched by {\it only\/} massive stars.  Most of the dwarf spheroidal
companions to the Milky Way have had extended star formation, and so
are expected to show evidence in the elemental abundances for the
incorporation of iron from long-lived Type~Ia supernovae. The Ursa
Minor dSph is the best candidate for having had a sufficiently short
duration of star formation that a significant fraction of its low-mass
stars formed prior to the onset of Type~Ia supernovae in sufficient
numbers to be noticed in the chemical elemental abundances; this
timescale is uncertain, but likely to be of the order of 1--2~Gyr. 

The elemental mix produced by  a generation of massive stars depends on the
massive-star mass function, because the yields of a given Type II
supernova depends on its mass. In particular, the $\alpha$-element
yields (nuclei formed by successive addition of a helium nucleus) 
vary more strongly with progenitor mass than does the iron
yield (e.g.~Figure~1 of Gibson 1998). There appears to have been
surprisingly good mixing at early times, at least in the stellar halo
of the Milky Way (see the remarkably low scatter in the ratio of
[$\alpha$/Fe] at [Fe/H] $\simlt -2.5$ in the sample analysed by
Cayrel et al.~2004), so that a well-defined value of [$\alpha$/Fe] is
produced by a generation of massive stars of given IMF.  This is seen
as the `Type~II plateau' in [$\alpha$/Fe] for metal-poor Galactic
stars. 

The available elemental abundance data for a handful of individual
stars in the UMi dSph are consistent with the same value for the
Type~II plateau as seen in stars of the Milky Way (Shetrone et
al.~2001), with some downturn for more metal-rich stars, as expected
if there is an age spread of 1--2~Gyr and an age-metallicity
relationship.  The simplest interpretation is that the high mass IMF
was the same in the UMi dSph as in the Galaxy -- and that IMF is a
power-law with Salpeter (1955) slope.

Most of the stars in the other dwarf spheroidals have
low values of [$\alpha$/Fe], consistent with a standard -- Salpeter --
IMF for massive stars and an extended star formation history, as implied by their
colour-magnitude diagram (see e.g.~Venn et al.~2004).

\section {Conclusions}

The fossil record in low-mass stars at the present time allows the
derivation of the stellar IMF at high redshift.  The low-mass
luminosity 
function is accessible through star counts, most robustly in a system
with a simple stellar population.  We have found that the low-mass IMF
is invariant between globular clusters in the halo of the Milky Way 
and an external galaxy, the dwarf spheroidal in Ursa Minor. The
underlying  mass
function is apparently the same as that for present-day star formation in the
local disk of the Milky Way.  The low-mass IMF is remarkably
invariant, over a broad range of metallicities, age, star-formation
rate, baryonic density, dark matter content -- indeed most of the
parameters that {\it a priori\/} one might have expected to be
important in determining the masses of stars.  The high-mass IMF is
also apparently independent of these  parameters.  This invariance is
particularly surprising if the Jeans mass plays an important role.

\begin{acknowledgments}
I  thank my colleagues and collaborators Sofia Feltzing, Jay
Gallagher, Gerry Gilmore, Leslie Hebb, Mark Houdashelt and Tammy
Smecker-Hane for their contributions to the results described here.
I would also like to thank the tireless organizers of this stimulating
meeting for inviting me.  
\end{acknowledgments}


\begin{chapthebibliography}{1}
\bibitem{} Baraffe, I., Chabrier, G., Allard, F. \& Hauschildt,
P. 1997, A\&A, 327, 1054
\bibitem{} Bullock, J., Kravtsov, A. \& Weinberg, D. 2000, ApJ, 539, 517

\bibitem{} Bellazzini, M., Ferraro, F., Origlia, L., Pancino, E.~et al.~2002, AJ, 124, 3222

\bibitem{}
Carrera, R., Aparicio, A., Martinez-Delgado, D. \& Alonso-Garcia,
J. 2002, AJ, 123, 3199
\bibitem{} Cayrel, R. et al.~2004, A\&A, 416, 1117
\bibitem{} Cole, S., Lacey, C., Baugh, C. \& Frenk, C. 2000, MNRAS,
319, 168
\bibitem{}Feltzing, S., Wyse, R.F.G. \& Gilmore, G. 1999, ApJL, 516, 17
\bibitem{} Gallagher, J.S. \& Wyse, R.F.G. 1994, PASP, 106, 1225
\bibitem{} Gibson, B. 1998, ApJ, 501, 675

\bibitem{} Hebb, L., Wyse, R.F.G. \& Gilmore, G. 2004, AJ, December
issue (astro-ph/0409289)
\bibitem{} Hernandez, X., Gilmore, G. \& Valls-Gabaud, D. 2000, MNRAS,
317, 831
\bibitem{} Kauffmann, G., White, S.D.M. \& Guiderdoni, B. 1993, MNRAS,
264, 201
\bibitem{} Kleyna, J., Geller, M., Kenyon, S., Kurtz, M. \&
Thorstensen, J. 1998, AJ,
115, 2359
\bibitem{} Palma, C., Majewski, S., et al.~2003, AJ, 125, 1352
\bibitem{} Piotto, G., Cool, A. \& King, I. 1997, AJ, 113, 1345
\bibitem{} Salpeter, E.E. 1955, ApJ, 121, 161
\bibitem{} Shetrone, M., C\^ot\'e, P. \& Sargent, W. 2001, ApJ, 548,
592
\bibitem{} Venn, K. et al.~2004, AJ, 128, 1177
\bibitem{} Wilkinson, M., Kleyna, J., Evans, W., Gilmore, G., Irwin,
M. \& Grebel, E. 2004, ApJL, 611, 21
\bibitem{} Wyse, R.F.G. 1998, in The Stellar IMF, ASP Conf series 142,
eds Gilmore \& Howell, p89
\bibitem{} Wyse, R.F.G. et al.~2002, New Ast, 7, 395

\end{chapthebibliography}
\end{document}